\input{psfig}
\documentstyle[sprocl]{article}

\def\be{\begin{equation}}
\def\ee{\end{equation}}
\def\bea{\begin{eqnarray}}
\def\eea{\end{eqnarray}}

\begin{document}

\title{Mixed Direct-Iterative Methods for Boundary Integral Formulations
of Dielectric Solvation Models}

\author{Steven A.  Corcelli, Joel D.  Kress, Lawrence R.  Pratt, Gregory
J.  Tawa}

\address{Theoretical Division, Los Alamos National Laboratory,Los
Alamos, NM 87545}


\maketitle\abstracts{This paper describes a mixed direct-iterative
method for boundary integral formulations of dielectric solvation
models.  We give an example for which a direct solution at thermal
accuracy is nontrivial and for which Gauss-Seidel iteration diverges in
rare but reproducible cases.  This difficulty is analyzed by obtaining
the eigenvalues and the spectral radius of the iteration matrix.  This
establishes that the nonconvergence is due to inaccuracies of the
asymptotic approximations for the matrix elements for accidentally close
boundary element pairs on different spheres.  This difficulty is cured
by checking for boundary element pairs closer than the typical spatial
extent of the boundary elements and for those pairs performing an
`in-line' Monte Carlo integration to evaluate the required matrix
elements.  This difficulty are not expected and have not been observed
when only a direct solution is sought.  Finally, we give an example
application of these methods to deprotonation of monosilicic acid in
water.}

\section{Introduction}

An interesting development in computational molecular biophysics over
the past decade has been the surprising utility of dielectric
models of solvation of molecular solutes in
water.\cite{rashin-review}$^{\!-\,}$\cite{bharadwaj} This is surprising
{\em a priori} because this approach neglects almost all of the
molecular characteristics of solvation.  {\em A posteriori} molecular
calculations have become available checking the basic soundness of the
dielectric model results\thinspace\cite{jayaram}$^{\!-\,}$\cite{kw} and
checking features of the underlying molecular
theory.\cite{lbk}$^{\!-\,}$\cite{hpgb}

Arguments that support such models are simple and broad:  much of the
solvation phenomena in water are dominated by electrostatic
interactions.  These models provide a physical description of solvation
of electrostatic interactions.  If we permit a macroscopic empirical
parameterization then they are indeed useful.  Furthermore, dielectric
models permit a conceptually natural and feasible coupling of solvation
theory with electronic structure tools of traditional computational
chemistry.  For these reasons too the dielectric models have been
helpful.\cite{estruc}

The numerical challenge in applying these models is the solution of the
Poisson equation \begin{equation} \nabla \bullet \varepsilon ({\bf
r})\nabla \Phi ({\bf r})=-\;4\pi \rho({\bf r}) \label{poisson}
\end{equation} where $\rho({\bf r})$ is the density of electric charge
associated with the solute mole\-cule, $\varepsilon({\bf r})$ gives the
local value of the dielectric constant, and $\Phi({\bf r})$ is the
electric potential.  This equation can be a challenge because the
function $\varepsilon({\bf r})$ changes abruptly on the modeled
molecular surface of the solute and that surface must sometimes exhibit
nontrivial variation on an atomic scale.

Because the most important difficulty is associated with treatment of
the molecular surface, boundary integral methods are
advantageous.\cite{yoon-lenhoff,bharadwaj} Those methods permit the
concentration of numerical resources on the description of the molecular
surface.  The resolution in the description of the molecular surface can
then be directly associated with the accuracy of the numerical
calculation.

The accuracy requirements of relevance to us are associated with
conformation free energy differences comparable to k$_B$T and with
treatment of the effects of molecular solvent structure by integrating
out probe water molecules with the help of this dielectric
model.\cite{pthgc} These interests put high demands of accuracy and
speed on the numerical methods.  Stringent testing of the accuracy of
these dielectric models for structural optimization has been pursued
only relatively recently.\cite{pthgc}

It might be questioned whether it makes sense to solve approximate
dielectric models to the accuracy discussed here.  We offer two
responses.  First, though the model is approximate, attempts to draw
conclusions from the model results are complicated by non-physical
errors superposed on the model results.  Second, if the model results
are valid enough to be helpful, then they might serve as an initial
approximation upon which more refined treatments might be
built.\cite{pthgc} In that case, understanding the accuracy of the
initial predictions would be important.

The accuracy that can be achieved in the solution of Eq.~(\ref{poisson})
through a direct boundary integral approach will be limited by the
dimension of the set of linear equations that corresponds to the linear
boundary integral equation.  Since the dimension of that set will be
relatively small if direct solution methods are used, substantial
numerical resources can be invested in obtaining accurate coefficients
for the linear equations.

When direct methods become unfeasible, it is natural to apply iterative
solution methods to the direct solution used as an initial estimate.
Because the initial solution is expected to be good, the iterative
effort is expected to be modest.  Iterative methods permit a larger
number of linear equations.  But numerical sophistication in the
evaluation of the larger number of matrix elements becomes prohibitive.
Thus, the price to be paid for the higher resolution is that the most
matrix elements are obtained `on the fly.'  The methodological problem
of this paper is the formulation of the mixed direct-iterative methods
for solution of Eq.~(\ref{poisson}); and the identification and
correction of a difficulty that can arise.

\begin{figure}
\hspace{0.8in}
\psfig{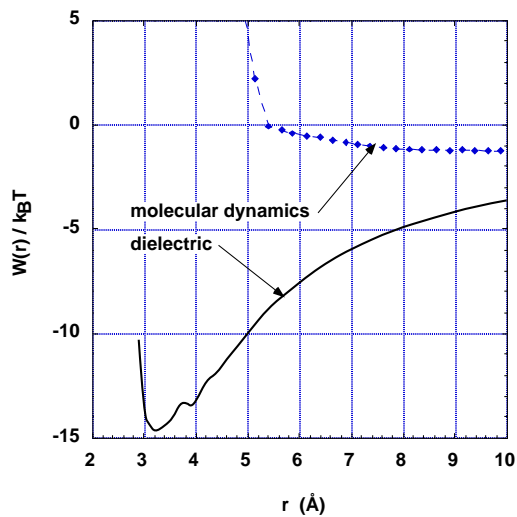}
\caption{Ca$^{++} \cdots $Cl$^-$ potential of mean force in water at
normal pressure and 298K.  The curve labeled `molecular dynamics' is the
literature molecular result.\protect\cite{guardia:93} The curve labeled
`dielectric' is obtained by the revised method of Section 2.4.
\label{cacl.fig}} \end{figure}

The results of Fig.~\ref{cacl.fig} present an example that will be used
because an iterative difficulty can be reproducibly exhibited.  Shown
there are calculations of the potential of mean force between a
Ca$^{++}$ ion and a Cl$^-$ ion in water.\cite{guardia:93} These results
utilize the van der Waals surface\thinspace\cite{pthgc} and the radii
recommended by Rashin and Honig.\cite{rashin85} We include here some
qualitative notes about physical aspects of these results.  Firstly, we
have much less experience with simulation results for this potential of
mean force than we do, for example, with Na$^+ \cdots$ Cl$^-$.  Thus, we
view the simulation results of Fig.~\ref{cacl.fig} as preliminary.
Assuming those results are born-out by further study, the interpretation
would be that the Ca$^{++}$ holds its solvation shell sufficiently
tightly that no contact minimum exists.\cite{guardia:93} Secondly, the
dielectric model result predicts an over deep contact minimum.  In this
respect the present results are consistent with previous
comparisons\thinspace\cite{rashin89,phg,tp,pthgc} and these results are
therefore not newly troubling.  Thirdly, the maximum in the dielectric
model result near 3.8\AA\ where the spheres just touch is expected to be
correct though it is clearly a subtle feature on the global scale shown
here.  The free energy of the separated ion pair is approximately 700
k$_B$T so resolution of such features here requires an relative accuracy
of about 0.1\%.  Still, the relative height of that maximum is not
negligible on a k$_B$T energy scale.  Calculations that establish the
correctness of k$_B$T features require
care.\cite{simonson,tucker,bharadwaj} Such methods are the topic of this
paper.

\section{Methods}

Here we catalog the methodological results used below for solution of
the Poisson Eq.~(\ref{poisson}).  Further discussion of the genesis of
these results can be found elsewhere.\cite{pthgc} We first cast
Eq.~(\ref{poisson}) as an integral equation, {\it e.g.\/}:
\begin{equation} \varepsilon({\bf r})\Phi({\bf r})= \Phi^{(0)}({\bf r})
+\int { \left[ {{\left( {{\bf r}-{\bf r}'} \right)\bullet \nabla'
\varepsilon ({\bf r}')} \over {4\pi \left| {{\bf r}-{\bf r}'}
\right|^3}} \right] {\Phi({\bf r}')} d^3r'} .  \label{inteq}
\end{equation} The quantity $\Phi^{(0)}({\bf r})$ is the electrostatic
potential in the absense of the medium.  Because the model assumes that
$\varepsilon({\bf r})$ has a sharp step at the molecular surface the
integration on the right collapses to a 2-dimensional integration over
the molecular surface.  That molecular surface is defined as the
boundary between the molecular volume --- modeled as the union of
spherical volumes centered on solute atoms --- and the solution region.
For ${\bf r}$ infinitismally outside the molecular surface,
Eq.~(\ref{inteq}) provides a closed equation for $\Phi({\bf r})$ on the
molecular surface.  Once $\Phi({\bf r})$ is obtained on the molecular
surface, it can be used on the right side of Eq.~(\ref{inteq}) to
construct the potential elsewhere.

{}From such solutions we construct the interaction part of the chemical
potential of the solute as \begin{equation} \Delta \mu^{(x)} =\left( {1
\over 2} \right)\int {\rho({\bf r})\left( {\Phi_l ({\bf r})-\Phi _v({\bf
r})} \right)d^3r}.  \label{x-mu} \end{equation} The subscripts $l$ and
$v$ indicate `liquid' and `vapor,' respectively, so that this difference
is the electric work required to charge the solute in the liquid
relative to the vapor.  This requires the solution of Eq.~(\ref{inteq})
twice, once for the liquid with \begin{equation} \varepsilon _l({\bf
r})=\varepsilon _m+\left( {\varepsilon _s-\varepsilon _m} \right)\eta
({\bf r}) , \label{eliq} \end{equation} and once for the vapor with
\begin{equation} \varepsilon _v({\bf r})=\varepsilon _m+\left(
{1-\varepsilon _m} \right)\eta ({\bf r}).  \label{evap} \end{equation}
Here $\eta ({\bf r})$ is a step function that is one outside the
molecular volume and zero otherwise; $\varepsilon_s$ is the dielectric
constant of the solution and $\varepsilon_m$ is an assigned `dielectric
constant of the molecule.'  The latter parameter is used to match the
polarizability of the solute.  The formulation Eq.~(\ref{inteq}) makes
it simple to match a given polarizability by adjustment of
$\varepsilon_m$.\cite{pthgc} This is because the kernel is proportional
to the electrostatic potential due to a surface dipole density.  Thus,
we perform calculations with $\varepsilon _v({\bf r})$ and with
$\Phi^{(0)}({\bf r})$ chosen to describe a uniform external electric
field.  The induced electrostatic potential in the far field is
associated with the induced dipole moment.  The correlation of the
induced dipole moment with the external field strength provides the
modeled molecular polarizability.

\subsection{Rules for coarse direct calculations}

A discretized version of Eq.~(\ref{inteq}) is \begin{equation}
\varepsilon_s \Phi({\bf r}_\alpha) = \Phi^{(0)}({\bf r}_\alpha) +
\sum_\beta w_{\alpha\beta}\Phi({\bf r}_\beta) .  \label{discrete}
\end{equation} Here {\bf r}$_\alpha$ is the $\alpha$-th `plaque point'
--- a point on the molecular surface obtained by a uniform sampling, for
example by exploiting either quasi-random number series or `good
lattice' procedures.\cite{hammersley}$^{\!-\,}$\cite{press} The plaques
are defined as the Voronoi polyhedra of the plaque points on the
nonburied surface of each sphere.  The matrix of coefficients
w$_{\alpha\beta}$ can be obtained as follows:  \begin{equation}
w_{\alpha \beta} = {R(s_\beta)^2 { (\varepsilon_s - \varepsilon_m)
}\over M(s_\beta)} \sum_{\{i\in \beta \}} { ( {\bf r}_\alpha - {\bf r}_i
) \bullet {\hat {\bf n} }_i \over |{\bf r}_\alpha - {\bf r}_i |^3 } ,
\qquad \alpha \ne \beta , \label{cod} \end{equation} and
\begin{equation} w_{\alpha\alpha} = {1 \over 2\sqrt 2}\left(
{\varepsilon_s - \varepsilon_m \over M(s_\alpha)}
\right)\sum\limits_{\left\{ {i\notin \alpha} \right\}} {\left( {1-\cos
\vartheta _{i\alpha}} \right)^{-1/2}} .  \label{cd} \end{equation} These
are Monte Carlo estimates of plaque integrations.  Further details can
be found elsewhere.\cite{pthgc} The set of sampling points that are
within the plaque $\beta$ is denoted by $\{i\in \beta\}$.  $M(s_\beta)$
is the number of points on the sphere s$_\beta$ that supports plaque
$\beta$.  In Eq.~(\ref{cd}), $\vartheta_{i\alpha}$ is the angle between
the surface normals at plaque point $\alpha$ and the sampled point $i$.
That formula arranges to use sample points {\it outside\/} the plaque to
calculate the solid angle subtended by that plaque at the plaque point.
The purpose is to reduce the variance of the Monte Carlo estimate.  All
sample points on the surface of the sphere that supports plaque $\alpha$
should be used in the estimation.  But whether any sample point resides
on plaque $\alpha$ depends on resolution of buried surface because the
plaque boundaries sometimes follow the boundaries between the exposed
and buried surface.

For accurate calculations on small molecule solutes, we have found that
the computational time is dominated by the Monte Carlo effort.  Thus, we
use these formulae differently to avoid some of that effort.  We use
Eq.~(\ref{cod}) with no Monte Carlo sample in addition to the plaque
point; these formulae constitute one point estimates then.  However, the
set of plaque points is now a larger set of equidistributed surface
points, elements of a fine lattice.  We form the equations to be
analyzed by contraction of that description to one based upon coarse
plaques constructed from an initial sequence of the fine lattice points.
We then require the equality of the potential at all fine lattice points
residing on the same coarse plaque.  This requirement results in an
overdetermined system.  This system is analyzed with a singular value
decomposition to obtain the plaque potentials that minimize the mean
square residual of those equations.\cite{press}

The advantages of this approach are that much of the Monte Carlo effort
can be avoided and that some account is taken of the spatial variation
of $\Phi^{(0)}({\bf r})$ within a coarse plaque.  The principal
disadvantage is that this approach requires more memory and this
disadvantage can be severe.

\subsection{Rules for Gauss-Seidel iteration}

Here we give the rules used in our iterative calculations.  We begin
with an approximate solution $\Phi({\bf r}_\alpha)$.  That approximation
is then updated in place according to \begin{equation} \Phi ({\bf
r}_\alpha )\leftarrow \left( {\varepsilon _s-w_{\alpha \alpha }}
\right)^{-1}\left\{ {\Phi ^{(0)}({\bf r}_\alpha )+\sum\limits_{\beta \ne
\alpha } {w_{\alpha \beta }\Phi ({\bf r}_\beta )}} \right\}
\label{gsrule} \end{equation} sequentially for all $\alpha$.  In this
calculation the plaque points are the points of the fine lattice.
Because that set of points is expected to be large, the off-diagonal
coefficients $w_{\alpha\beta}$ are evaluated `on the fly' using
Eq.~(\ref{cod}) but no Monte Carlo sample in addition to the
plaque point.  Our experience is that accurate evaluation of the
diagonal coefficients $w_{\alpha\alpha}$ is important, so we evaluate
them at an initial stage of the calculation and store them for later
use.

\begin{figure} \hspace{0.8in} \psfig{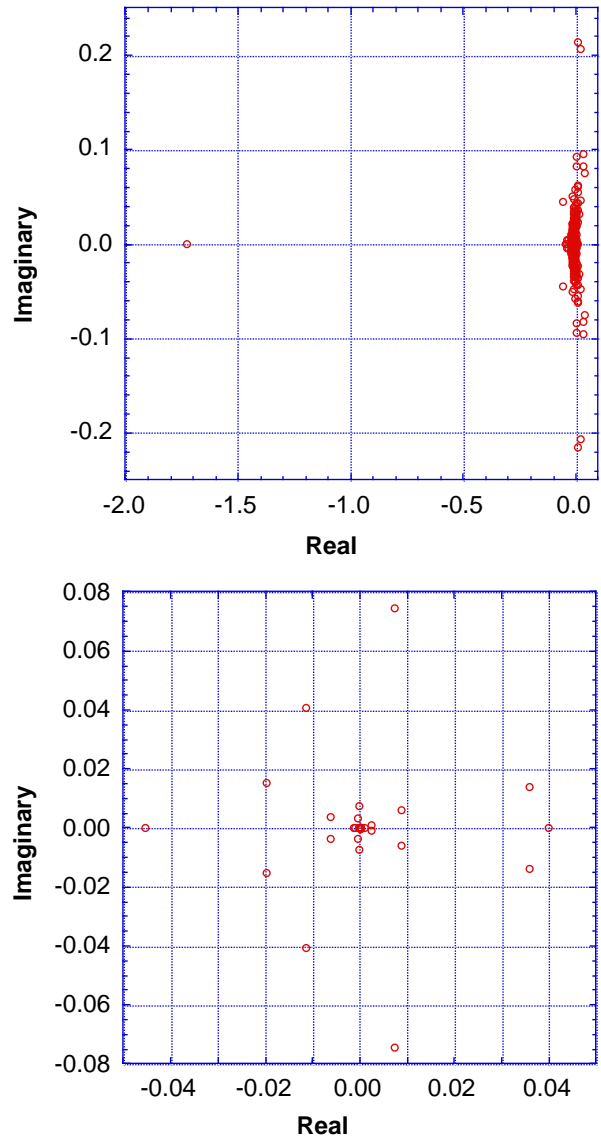}
\caption{Upper panel:  Eigenvalues of the Gauss-Siedel iteration matrix
obtained as described in Section 2.2 for 36 plaques on the Ca$^{++}
\cdots$ Cl$^-$ di-ion for r = 3.7\AA.  One eigenvalue is much further
from the origin than 1.0.  Lower panel:  Eigenvalues of the Gauss-Siedel
iteration matrix obtained by the revised approach described in Section
2.3.  \label{srad.fig}} \end{figure}

These methods were applied to the calculation of the pair potential of
the mean forces between a Ca$^{++}$ and a Cl$^{-}$ in water.  It was
found that the Gauss-Seidel iteration scheme converged almost always,
but diverged in rare but reproducible circumstances.  A necessary and
sufficient condition for the convergence of Gauss-Seidel iteration is
that the spectral radius of the iteration matrix must be less than
one.\cite{strikwerda} Plotted in Fig.~\ref{srad.fig} are the eigenvalues
of the Gauss-Seidel iteration matrix for the Ca$^{++}\cdots $ Cl$^{-}$
problem for the case of $r$ = 3.7\AA\ with 36 plaque points.  One
eigenvalue is much greater than one.  Also plotted there are eigenvalues
of the Gauss-Seidel iteration matrix for the same circumstances except
that matrix elements were obtained by a modified method described below.
All eigenvalues are now substantially less than one.  Using the
corrected methods the divergences have not been observed.

\subsection{What we did to fix the problem}

It was when those Monte Carlo efforts were economized that iterative
divergence occasionally presented a problem.  Furthermore, the diagonal
matrix elements are always calculated the same way.  This suggests that
the observed difficulty was due to the one-point estimate of the
off-diagonal elements used when the iterative calculation was
implemented.

The estimate Eq.~(\ref{cod}) will have a larger variance the closer the
point ${\bf r}_\alpha$ to plaque $\beta$.  Thus, it is reasonable to
suspect those matrix elements corresponding to close $\alpha\beta$.  It
was verified that this suspicion is correct by replacing
$w_{\alpha\beta}$ by $w_{\beta\beta}$ whenever ${\bf r}_\alpha$ is on a
different sphere than plaque $\beta$ and $\vert{\bf r}_\alpha-{\bf
r}_\beta\vert < 2 R(s_\beta)/\sqrt{M(s_\beta)} $.  This is a statistical
estimate of the radial extent of plaques on center s$_\beta$.  This
unsatisfying maneuver eliminated the divergence.

A geniune solution is to implement a Monte Carlo calculation of those
matrix elements $w_{\alpha\beta}$ identified as potentially problematic.
For ${\bf r}_\alpha$ close to plaque $\beta$, an approach like that of
Eq.~(\ref{cd}) using points sampled outside the plaque would be most
appropriate.  But for ${\bf r}_\alpha$ far from plaque $\beta$, it would
be natural to use points on the plaque.  That we could use either
approach when the point ${\bf r}_\alpha$ is not on the plaque is
justified by the relation \begin{equation} \int { \left[ {{\left( {{\bf
r}-{\bf r}'} \right)\bullet \nabla' \varepsilon ({\bf r}')} \over {4\pi
\left| {{\bf r}-{\bf r}'} \right|^3}} \right] d^3r'} =0, \label{glaw}
\end{equation} valid for {\bf r} outside the sphere.  This is an
application of Gauss's law.  Thus we could estimate the required
integral using either points on the plaque $\beta$ or on a complementary
spherical surface.  Using 1/2 of each estimate is an example of the
method of antithetic variates:\thinspace\cite{hh} \begin{equation}
w_{\alpha \beta} = {R(s_\beta)^2 { (\varepsilon_s - \varepsilon_m)
}\over 2M(s_\beta)} \left[ \sum_{\{i\in \beta \}} { ( {\bf r}_\alpha -
{\bf r}_i ) \bullet {\hat {\bf n} }_i \over |{\bf r}_\alpha - {\bf r}_i
|^3 } - \sum_{\{i\notin \beta \}} { ( {\bf r}_\alpha - {\bf r}_i )
\bullet {\hat {\bf n} }_i \over |{\bf r}_\alpha - {\bf r}_i |^3 }
\right].  \label{anit-check} \end{equation} This effort is expended only
when ${\bf r}_\alpha$ is on a different sphere than plaque $\beta$ and
$\vert{\bf r}_\alpha-{\bf r}_\beta\vert < 2 R(s_\beta)/\sqrt{M(s_\beta)}
$.

\section{Deprotonation of monosilicic acid in water}

As an application of the methods developed in the previous sections we
will treat the deprotonation of monosilicic acid in water at 298K
\begin{equation} Si(OH)_{4} \rightleftharpoons Si(OH)_{3}O^{-} + H^{+} .
\label{rx} \end{equation} The monosilicic acid molecule and anion are
depicted in Fig.~\ref{o-sili.fig}.  The equilibrium ratio is
\begin{equation} K={{\left[ {Si(OH)_3O^-} \right]\left[ {H^+} \right]}
\over {\left[ {Si(OH)_4} \right]}} \label{k} \end{equation} with
concentrations in molar units.  The quantity we seek is the free energy
of reaction \begin{equation} \Delta G^{(0)} \equiv -RT \ln K \label{A}
\end{equation} measured to be
$13.5$~kcal/mol.\cite{garrels}$^{\!-\,}$\cite{rustad}

This is a helpful example for several reasons.  Acid-base equilibria are
an important application of these models in molecular biophysics.
Additionally, these solutes have not be treated previously by these
methods.  Thus, the expectations for the radii-parameters required can
be tested outside the conventional parameterization suite of solutes.

\subsection{Solution thermodynamic formulation}

It is more physical\thinspace\cite{kw} to consider the reaction
\begin{equation} Si(OH)_{4} + H_{2}O \rightleftharpoons Si(OH)_{3}O^{-}
+ H_{3}O^{+} .  \label{rx1} \end{equation} The reaction described this
way does not result in a net loss of chemical bonds and this is likely
the case in water.\cite{kw} The ratio \begin{equation} \tilde K={
{\left[ Si(OH)_3O^- \right]\left[ H_3O^+ \right]} \over {\left[ Si(OH)_4
\right]\left[ H_2O \right]} } \label{ktilde} \end{equation} is then
dimensionless.  This equilibrium ratio may be obtained
as\thinspace\cite{kw} \begin{equation} \tilde K(T) = \tilde K^{(0)}(T)
\exp \left[-\Delta\Delta\mu ^{(x)} / RT \right] , \label{stat-mech}
\end{equation} where $\tilde K^{(0)}(T)$ is the ideal gas result
obtainable from standard formulae,\cite{McQuarrie} and \begin{equation}
\Delta\Delta\mu ^{(x)} \equiv \Delta \mu ^{(x)} _{H_3O^+}+\Delta \mu
^{(x)} _{Si(OH)_3O^-}-\Delta \mu ^{(x)} _{H_2O}-\Delta \mu ^{(x)}
_{Si(OH)_4} .  \label{rmp} \end{equation} $K$ and $\tilde K$ are related
by $ \tilde K={K / {\left[ {H_2O} \right]}}$ and the free energy of
reaction is simply \begin{equation} \Delta G^{(0)}=-RT \ln \tilde K - RT
\ln [H_{2}O] .  \label {Afinal} \end{equation} We assume that the solute
concentrations are sufficiently low that the formal concentration of
H$_2$O is satisfactory.

\subsection{Electronic structure results on the isolated molecules}

\begin{table}
\caption{Partial charges for Si(OH)$_4$ and
Si(OH)$_3$O$^-$.  \label{tab:scharges}}
\vspace{0.4cm}
\begin{center}
\begin{tabular}{|c|c|c|} \hline
Atom & Si(OH)$_4$ & Si(OH)$_3$O$^-$ \\ \hline
Si & 1.62 & 1.52 \\ \hline
O1 & -0.90 & -0.89 \\ \hline
O2 & -0.90 & -0.90 \\ \hline
O3 & -0.90 & -0.92 \\ \hline
O4 & -0.90 & -1.06 \\ \hline
H1 & 0.49 & 0.44 \\ \hline
H2 & 0.49 & 0.41 \\ \hline
H3 & 0.49 & 0.41 \\ \hline
H4 & 0.50 & --- \\ \hline
\end{tabular}
\end{center} \end{table}

\begin{figure}
\hspace{0.9in} \psfig{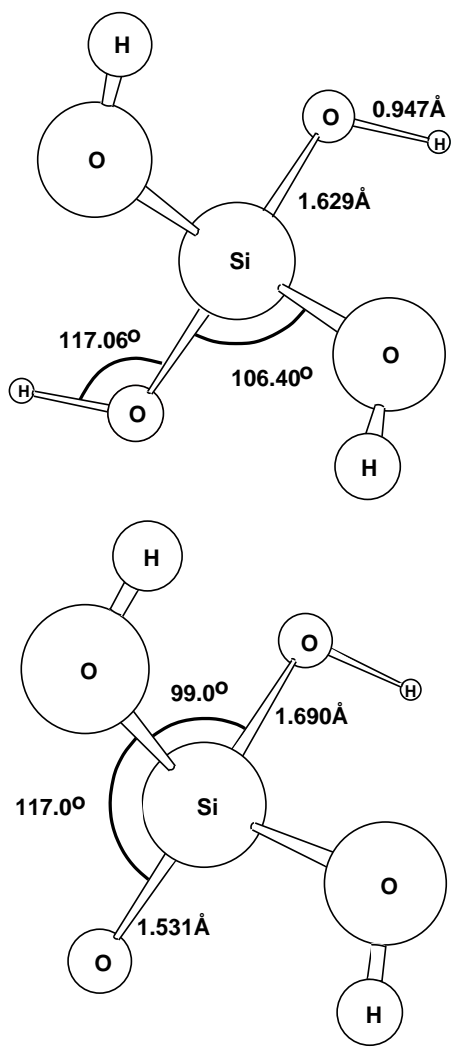}
\caption{Monosilicic acid molecule (upper) and anion (lower)
established by the electronic structure calculations of Section 3.2
\label{o-sili.fig} } \end{figure}

All electronic structure calculations were performed using the
GAUSSIAN-92 program.\cite{Frisch:92} Two different quantum mechanical
methods were employed:  Hartree-Fock (HF), and HF followed by a
second-order order M{\"o}ller-Plesset (MP2) correlation energy
correction.  Two different basis sets were used:  6-31G(d) [also denoted
6-31G$^*$], and 6-31G++(2d).  The ``(d)'' and ``(2d)'' denotes that the
6-31G basis is supplemented by one and two sets of
polarization\thinspace\cite{Frisch:84} d-functions, respectively, on the
heavy (non-hydrogen) atoms.  The "++" denotes that the basis is
supplemented by diffuse\thinspace\cite{Clark} functions.  Together the
quantum mechanical method and basis set specifies a theoretical model,
{\it e.g.}, Sauer\thinspace\cite{Sauer:87} has performed HF/6-31G(d)
calculations on monosilicic acid.  The optimized geometries for H$_2$O,
H$_3$O$^+$, Si(OH)$_4$ and Si(OH)$_3$O$^-$ were determined by analytic
gradient techniques using the HF/6-31G(d) model.  The structures found
for monosilicic acid and its anion are shown in Fig.~\ref{o-sili.fig}.
The bond distance and angle for H$_2$O is 0.947\AA\ and
105.50$^{\circ}$, respectively compared to the experimental
values\thinspace\cite{herzberg:66} of 0.957\AA\ and 104.5$^{\circ}$,
respectively.  The calculated value for all three H-O-H bond angles for
H$_3$O$^+$ is 113.06$^{\circ}$.  Teppen {\it et al.\/}\cite{Teppen:94}
have examined the effects of basis set size and electron correlation
corrections on the properties of monosilicic acid.  For a 6-31G(d)
basis, they\thinspace\cite{Teppen:94} find that the MP2 bond lengths for
Si-O and O-H were 0.022 and 0.023\AA\ larger, respectively than the HF
values and the MP2 Si-O-H angle decreased 2.9$^{\circ}$ from the HF
value.  For the HF method, they\thinspace\cite{Teppen:94} also find that
the MC6-311G(2d,2p)\thinspace\cite{McLean:80} bond lengths for Si-O and
O-H were 0.007 and 0.009\AA\ smaller than the 6-31G(d) values and the
MC6-311G(2d,2p) Si-O-H bond angle increased 1.3$^{\circ}$ from the
6-31G(d) value.  For the present work these differences are acceptable,
and HF/6-31G(d) was used to optimize geometries.

\begin{table}
\caption{Partial charges for H$_2$O and H$_3$O$^+$. \label{tab:wcharges}}
\vspace{0.4cm}
\begin{center}
\begin{tabular}{|c|c|c|} \hline
Atom & H$_{2}$O & H$_{3}$O$^{+}$ \\ \hline
O & -0.82 & -0.62 \\  \hline
H & 0.41 & 0.54 \\ \hline
\end{tabular} \end{center} \end{table}

To obtain atom-centered charges (Tables~\ref{tab:scharges} and
\ref{tab:wcharges}) necessary for the solvation energy calculation, a
fit of the HF/6-31G(d) electrostatic potential
(CHELPG\thinspace\cite{breneman}) was performed.  Harmonic vibrational
frequencies and rotational constants were computed with the HF/6-31G(d)
model and were used to compute the partition functions in
Eq.~(\ref{stat-mech}).  Since the HF method overestimates frequencies,
the computed values were scaled by 0.88.\cite{hehre} The electronic
ground state energies, E$_0$, (Table~\ref{tab:ee}) were calculated by
optimizing the molecules with the MP2/6-31G(d) model.  The calculation
of polarizability, $\bar
\alpha=(\alpha_{xx}+\alpha_{yy}+\alpha_{zz})/3$, is sensitive to the
basis set.  For H$_2$O at the HF/6-31G(d) geometry, $\bar \alpha=0.70$,
0.90, 0.90, and, 1.06\AA$^3$, calculated with the 6-31G(d), 6-31+G(d),
6-31G(2d), and 6-31++G(2d) basis sets, respectively.  The 6-31++G(2d)
value agrees reasonably well with another HF
calculation,\cite{arrighini} $\bar \alpha=$ 1.17\AA$^3$.  The
experimental value\thinspace\cite{stillinger,moskowitz} for H$_2$O,
$\bar \alpha=$ 1.44\AA$^3$, was used in the solvation calculations for
both H$_2$O and H$_3$O$^+$.  For Si(OH)$_4$ and Si(OH)$_3$O$^-$, $\bar
\alpha$ was calculated with the HF/6-31G++(2d) model using the
HF/6-31G(d) geometries.  These values were scaled by 1.36, the ratio of
the experimental and HF/6-31G++(2d) $\bar \alpha$ values for H$_2$O.
The scaled values (Table~\ref{tab:ee}) were then used in the solvation
calculations.

\begin{table}
\caption{Electronic energies and polarizabilities. \label{tab:ee}}
\vspace{0.4cm}
\begin{center}
\begin{tabular}{|c|c|c|} \hline
Molecule & E$_0$ (hartree) & $\overline\alpha$ (\AA$^3$)\\ \hline
Si(OH)$_4$ & -590.89169 & 6.9 \\ \hline
Si(OH)$_3$O$^-$ & -590.29838 & 7.4 \\ \hline
H$_2$O &-76.01075 & 1.44 \\ \hline
H$_3$O$^+$&-76.28934 & 1.44 \\ \hline
\end{tabular} \end{center} \end{table}

\subsection{Results for the deprotonation of monosilicic acid}

     Three different calculations were performed for the free energy of
deprotonation, Eq.~(\ref{rx}).  The first calculation did not include
molecular polarizability, nor spheres on the H atoms of Si(OH)$_{4}$ and
Si(OH)$_{3}$O$^{-}$.  The second calculation included molecular
polarizability, but again not spheres on the H atoms.  The final
calculation included both molecular polarizability and spheres
on every atom of Si(OH)$_{4}$ and Si(OH)$_{3}$O$^{-}$.  This sequence of
calculations reflects our chronological approach to this system starting
with the simplest model, gradually adding more complications, and
gradually refining the values of the parameters used.  This approach
also gives helpful information on the sensitivity of the calculation to
the empirical parameters used.

The evaluation of the vibrational, rotational, and translational
partition functions of the isolated molecules on the basis of the
electronic structure results leads to a multiplicative contribution of
1.30 to the equilibrium ratio of Eq.~(\ref{stat-mech}).  The change in
the electronic energy $\Delta E_0$ was found to be 197.5~kcal/mol.

In all of the calculations the water and hydronium ions were treated as
single spheres of radius 1.6\AA\ on the O atom.  A molecular dielectric
constant $\varepsilon _{m}$ was 2.42.  This reproduces the
polarizability of the water molecule.\cite{kw,stillinger,moskowitz} The
solution dielectric constant was $\varepsilon _{s}=77.4$ appropriate to
water at 298K and 1.0g/cm$^3$.  The calculation of the excess chemical
potential of solvent species used 186 coarse lattice points and 936 fine
lattice points.  Ten Gauss-Seidel iteration passes were applied to the
coarse solution in all calculations.  The difference in excess chemical
potential between the hydronium ion and water was found to be
-107~kcal/mol.  All calculations on Si(OH)$_{4}$ and Si(OH)$_{3}$O$^{-}$
used 36 coarse lattice points and 936 fine lattice points on each
sphere.

The first calculations for Si(OH)$_{4}$ and Si(OH)$_{3}$O$^{-}$ used a
sphere of radius 1.8\AA\ on the each Si atom and a sphere of radius
1.65\AA\ on each O atom.  $\varepsilon _{m}=$1.0 was adopted, thus
ignoring the polarizability of the molecule.  The difference in excess
chemical potential was found to be -49.6~kcal/mol.  This calculation
gives 38.3~kcal/mol for the change in free energy for Eq.~(\ref{rx}), in
poor agreement with experiment.

In the next calculation a sphere of radius 1.8\AA\ was centered on the
each Si atom and a sphere of radius 1.60\AA\ on each O atom.
$\varepsilon _{m}=$2.95 and 3.40 were assigned to the Si(OH)$_{4}$ and
Si(OH)$_{3}$O$^{-}$ molecules, respectively.  These values match the
estimated molecular polarizabilities discussed in the previous section.
The difference in excess chemical potential was found to be -57.9
kcal/mol, which gives 30.0~kcal/mol for the change in free energy for
Eq.~(\ref{rx}).  This improves upon the calculation which ignored the
polarizability of Si(OH)$_{4}$ and Si(OH)$_{3}$O$^{-}$ but still differs
from experiment by more than a factor of two.

The final calculation had spheres on every atom of the solute
molecules.  The O atoms were given radii of 1.4\AA,
the Si atoms were given radii of 1.8\AA, and the H atoms were given
radii of 1.3\AA.  Values of 3.10 and 3.55 were used for the $\varepsilon
_{m}$ of Si(OH)$_{4}$ and Si(OH)$_{3}$O$^{-}$, respectively.  Again
these values were determined to match the polarizability of the
molecules.  The difference in excess chemical potential between
Si(OH)$_{4}$ and Si(OH)$_{3}$O$^{-}$ was -68.1~kcal/mol.  This
calculation gives 19.8~kcal/mol for the change in free energy for the
deprotonation of Si(OH)$_{4}$ in water, a much improved agreement with
experiment.

\section{Conclusions}

The iterative divergence occasionally encountered in the calculation of
the Ca$^{++} \cdots$ Cl$^-$ pair potential of mean force was due to
inaccuracies of the asymptotic approximations used for the matrix
elements for accidentally close boundary elements on different atomic
spheres.  This problem is cured by checking for boundary element pairs
closer than the typical spatial extent of the boundary elements and for
those pairs performing an `in-line' Monte Carlo integration to evaluate
the required matrix elements.  These difficulties are not expected and
have not been observed when only a direct solution is considered.

These methods can give a reasonable description of the free energetics
of the deprotonation of monosilicic acid in water.  A modeled solute
polarizability and spheres on hydroxyl protons have been found to be
important in achieving a reasonable agreement between model and
experiment.  The discrepancy remaining provides a suggestion of more
specific solute-solvent interaactions.  It has been noted
previously\thinspace\cite{pthgc} how specific molecular solvation
structure can be reintroduced into these models.  Those ideas for
integrating-out solvent degrees of freedom suggest the electronic
structure calculations should be performed on complexes of the solutes
of interest {\it plus\/} a probe water molecule.\cite{pthgc} Those
approaches will require a substantially larger computational effort.

The necessity of better treatment of the molecular solvation structure
is also clear in the example of Ca$^{++} \cdots$ Cl$^-$.  The probe
water molecule approach mentioned above would help here too but would
require accurate, rapid calculations on larger, more complicated
solution complexes.  It is hoped that the methods developed here will
make such calculations feasible.

\section*{Acknowledgements}

SAC acknowledges the support of Associated Western Universities.  We
thank C.  O.  Grigsby, P.  J.  Hay, and R.  L.  Martin for helpful
comments.

\end{document}